%% file: morales06.tex
\documentclass[preprint2]{aastex}

\shorttitle{shorttitle}
\shortauthors{shortauthor}

\begin{document}



\title{A Sensitive Search for Variability in Late L Dwarfs: The Quest for Weather}


\author{M.~Morales-Calder\'on\altaffilmark{1,}\altaffilmark{2}, J.R.~Stauffer\altaffilmark{3}, J.Davy~Kirkpatrick\altaffilmark{4}, S.~Carey\altaffilmark{3}, C.R.~Gelino\altaffilmark{3}, D.~Barrado~y~Navascu\'es\altaffilmark{1}, L.~Rebull\altaffilmark{3}, P.~Lowrance\altaffilmark{3}, M.S.~Marley\altaffilmark{5}, D.~Charbonneau\altaffilmark{6,7}, B.M.~Patten\altaffilmark{6}, S.T. Megeath\altaffilmark{6}, D.~Buzasi\altaffilmark{8} }

\email{mariamc@laeff.inta.es}


\altaffiltext{1}{Laboratorio de Astrof\'{\i}sica Espacial y F\'{\i}sica Fundamental (LAEFF), INTA. P.O.50727, E-28080 Madrid, Spain}
\altaffiltext{2}{Visiting Graduate Student Fellow at the Spitzer Science Center, California Institute of Technology, Pasadena, CA 91125 }
\altaffiltext{3}{Spitzer Science Center, California Institute of Technology, Pasadena, CA 91125}
\altaffiltext{4}{Infrared Processing and Analysis Center, California Institute of Technology, Pasadena, CA 91125}
\altaffiltext{5}{NASA Ames Research Center, Moffett Field, CA 94035}
\altaffiltext{6}{Harvard-Smithsonian Center for Astrophysics, Cambridge, MA 02138}
\altaffiltext{7}{Alfred P. Sloan Research Fellow}
\altaffiltext{8}{US Air Force Academy, Colorado Springs, CO 80840}


\begin{abstract}
  We have conducted a photometric monitoring program of 3 field late-L
  brown dwarfs (DENIS-P~J0255-4700, 2MASS~J0908+5032 and
  2MASS~J2244+2043) looking for evidence of non-axisymmetric structure
  or temporal variability in their photospheres.  The observations
  were performed using $Spitzer$/IRAC 4.5~$\mu$m and 8~$\mu$m
  bandpasses and were designed to cover at least one rotational period
  of each object. One-sigma RMS (root mean squared) uncertainties of
  less than 3~mmag at 4.5~$\mu$m and around 9~mmag at 8~$\mu$m were
  achieved.  Two out of the three objects studied exhibit some
  modulation in their light curves at 4.5~$\mu$m -- but not 8~$\mu$m
  -- with periods of 7.4~hr (DENIS0255) and 4.6~hr (2MA2244) and
  peak-to-peak amplitudes of 10~mmag and 8~mmag.  Although the lack of
  detectable 8~$\mu$m variation suggests an instrumental origin for
  the detected variations, the data may nevertheless still be
  consistent with intrinsic variability since the shorter wavelength
  IRAC bandpasses probe more deeply into late L dwarf atmospheres than
  the longer wavelengths.  A cloud feature occupying a small
  percentage (1-2~\%)of the visible hemisphere could account for the
  observed amplitude of variation.  If, instead, the variability is
  indeed instrumental in origin, then our non-variable L dwarfs could
  be either completely covered with clouds or objects whose clouds are
  smaller and uniformly distributed. Such scenarios would lead to very
  small photometric variations. Followup IRAC photometry at 3.6~$\mu$m
  and 5.8~$\mu$m bandpasses should distinguish between the two
  cases. In any event, the present observations provide the most
  sensitive search to date for structure in the photospheres of late-L
  dwarfs at mid-IR wavelengths, and our photometry provides stringent
  upper limits to the extent to which the photospheres of these
  transition L dwarfs are structured.
\end{abstract}




\section{Introduction}

The onslaught of L and T dwarf discoveries within in the last ten
years has enabled direct comparisons between observations and modeling
of brown dwarf cooling tracks.  The transition region from the late L
dwarfs to the early T dwarfs has always been problematic for brown
dwarf atmosphere modelers.  First among the unanswered questions
relates to the fact that early T dwarfs tend to have absolute $J$
magnitudes brighter than later L dwarfs, the so-called $J$-band
``bump'' (Vrba, Henden, Luginbuhl et al. 2004).  Further problems
arise from the large dispersion of certain colors as a function of
spectral type (Knapp, Legget, Fan et al. 2004) as well as the
discrepancies between the optical and near-IR derived spectral types
of some transition objects
\citep{Kirkpatrick05}.  Though some of these issues can be answered
with unresolved binaries \citep{Liu06}, it is also likely true that
the mechanism for dust clearing is intimately involved in the
explanation of all of these observables.  At least three mechanisms
for dust clearing have been proposed: (a) the cloud deck thins and
sinks, eventually dropping below the photosphere \citep{Tsuji03}; (b)
the cloud deck breaks up into discrete (patchy) clouds, and eventually
those clouds either shrink or sink below the visible photosphere
(Burgasser, Marley, Ackerman et al. 2002); and (c) a ``sudden
downpour'' (rapid sedimentation period) occurs, rapidly removing
grains from the visible photosphere \citep{Knapp04}.

Photometric variability is one observable that may be able to provide
constraints on which of these mechanisms, if any, is the dominant
process occurring very cool atmospheres.  The atmospheres of
these objects are too cool and neutral to support star spots (Mohanty
\& Basri 2003; Gelino, Marley, Holtzman et al. 2002), so if
variability exists, it is most likely caused by non-uniform structures
in the cloud deck.  If the object are not variable, then either the
variability is below the limits of detection, or the cloud decks are
uniformly distributed over the entire atmosphere, leaving no features
to produce brightness variations.

Numerous attempts have been made to search for photometric variability
in L and T dwarfs.  These searches for ``weather'' have been performed
largely in the optical regime (Tinney \& Tolley 1999; Bailer-Jones \&
Mundt 1999, 2001; Clarke, Oppenheimer, \& Tinney 2002a; Clarke, Tinney
\& Covey 2002b; Gelino 2002; Gelino et al.2002; Koen 2003; Koen 2005;
Maiti, Sengupta, Parihar et al. 2005) and the near-IR
\citep*{Bailer02,GelinoPhD02,Bailer03,Enoch03,Koen04,Koenetal05}.  The
results of these surveys indicate that the photometric variability of
these objects falls under one of three categories: non-variable,
periodic variable, and non-periodic variable.  Objects that show no
variations generally have limits of a few percent.  Those that show
non-periodic variations have RMS (root mean squared) amplitudes of a
few percent and vary on timescales too short to be correlated with a
rotation period \citep{Bailer01,Gelino02,Bailer04}.  The small
fraction that appear to show periodic modulation of their light curves
have typical amplitudes of a few percent and periods usually of order
several hours. The fact that the light curves in several cases appear
roughly sinusoidal suggests high latitude features; low latitude
features (i.e. those near the equator) would likely be eclipsed when
on the far side of the object, resulting in a flat section in the
light curve, and this is not observed\footnote{\citet{Gelino02}
observed a significant dip occurring over 10 days in the generally
flat light curve of the L1 dwarf 2MASS 1300+1912.  The duration of
this feature is much longer than the expected rotation period of an L
dwarf (a few hours; (Basri, Mohanty, Allard et al. (2000);
Bailer-Jones (2004)) and possibly reflects the creation and subsequent
dissipation of a large storm.}. Another explanation would be that the
clouds are distributed on the surface so that one hemisphere is
cloudier than the other, that is, we are seeing changes in the cloud
covering as the brown dwarf rotates.

The limiting factors in the photometric accuracy of these surveys are
the intrinsic faintness of the targets in the optical and second order
extinction effects from the Earth's atmosphere in the near-IR
\citep{Bailer03}.  In both cases, the usual single measurement
one-sigma uncertainties is of order the amplitude of the quoted
variability. This effect could be responsible for objects appearing
variable in one survey, but not in others \citep{Bailer01,Gelino02}.
It is also possible that some claims of variability in L dwarfs are
spurious and instead are the result of higher-than-expected
photometric errors. Only highly precise photometric observations can
resolve such issues.

We have conducted a program with the InfraRed Array Camera (IRAC) on
the $Spitzer$ Space Telescope to search for photometric variations in a
small sample of late L dwarfs near the L/T transition. In the next
section we describe briefly the target selection. Section~3 describes
the observational strategy used to accomplish the desired accuracy
level. Sections~4 and 5 deal with the data reduction and the
correction of the instrumental effects found in our data and section~6
describes briefly the variability and periodogram tests that we have
used. We present the results for each target in section~7 and,
finally, summarize our findings in section~8.


\section{Target Selection}

Our sample consists of 3 late L field brown dwarfs near the L/T
transition. Two of them were selected based on having a large
$v\sin{i}$, and hence a period easy to cover with a few hours of
continuous monitoring, and the third was selected based on NIR colors.
The principal characteristics of these objects are shown in
Table~\ref{targets} and their IRAC magnitudes in Table~\ref{results}.

DENIS-P J0255-4700 (hereafter DENIS0255) is an L8 brown dwarf
\citep{Kirkpatrickinprep} at approximately 5.0 pc.  It is one of the
brightest members of the so-called late-L/early-T ``transition''
objects that are the subject of this work. This object has a
$v\sin{i}$ of 40~$\pm$~10~km~s$^{-1}$ measured by \citep{Basri00} and
$v\sin{i}$ of 40.8~$\pm$~8.0~km~s$^{-1}$ or 41.1~$\pm$~2.8~km~s$^{-1}$
measured by \citep{Zapatero06}. For an object of radius 0.1~$R_\odot$,
as expected for brown dwarfs
\citep{Chabrier97}, a rotational velocity of 40~km~s$^{-1}$
corresponds to a rotation period of 3 hours; this is an upper
limit due to the unknown inclination of the rotation axis to the line
of sight.

The second object, 2MASS J0908+5032 (hereafter 2MA0908) is one of a
handful of L dwarfs with very discrepant optical and near-infrared
spectral types. In the optical its type is L5 (Cruz, Reid, Liebert et
al. 2003), but its near-infrared type is much later, L9 $\pm$ 1
\citep{Knapp04}.  The late near-infrared type could be an indicative
of a cloudy atmosphere, while the optical type indicates a temperature
warmer than the average, a very dusty dwarf.  This object has a
measured $v\sin{i}$ of 31~km~s$^{-1}$ (D.~Charbonneau, personal
communication 2006), so its period should be less than 4 hours.

The last object, 2MASS J2244+2043 (hereafter 2MA2244), is a brown
dwarf with a spectral type of L7.5~$\pm$~2 \citep{Knapp04}. Although
its $v\sin{i}$ has not been measured, it was selected as a target
because it is among the reddest known L or T dwarfs in the
near-infrared colors ($J-K_s = 2.45$). As such, it is believed to be
exceedingly dusty and thus a prime candidate for this work. Based on
the average $v\sin{i}$ (in the range 20-40~km~s$^{-1}$) of L
dwarfs of similar type \citep{Basri00, Mohanty03, Bailer04,
Zapatero06}, we expect a rotational period of approximately 6.5 hours
or less.

\placetable{targets}


\section{Observations}

The goal of our program was to obtain well-sampled relative photometry for our
target objects for time periods longer than their expected rotational period.
We hoped to be able to do both temporal relative photometry (i.e., how the
measured flux of our target objects varied with time during the observation)
and differential relative photometry (i.e., how the brightness of our targets
varied as compared to another comparison object in our field of view).  In
general, the comparison stars we had hoped to use proved to be fainter than
expected, making their photometry less accurate, and for that reason most of
the results we report will simply be for the temporal relative photometry of
the brown dwarf itself.

IRAC has four separate cameras, and data are collected in all four
channels (3.6, 4.5, 5.8 and 8.0 $\micron$) for the standard AOR
(Astronomical Observation Request).  The detector arrays have been
shown to be very stable, with very little variation in the flux
calibration over the entire time $Spitzer$ has been in orbit (Fazio,
Hora, Allen et al. 2004; Reach, Megeath, Cohen et al. 2005).  IRAC is
also very sensitive, and is capable of obtaining enough photons for
milli-magnitude photometry in at least Channels 1 and 2 for all of our
targets with integration times of 100 seconds or less.  Given these
expectations, the primary limitations for temporal relative photometry
would come from flat-field errors and other pixel-to-pixel effects.
This suggests that an observing mode where the spacecraft simply
stares at the target object, with no dithering, should provide the
most accurate relative photometry.  This expectation has been
confirmed by the recent usage of IRAC to measure the depth of the
planet transit in TrES-1 \citep{Charbonneau05}, where relative
photometry with RMS accuracies of 0.5 and 1.5 mmag were demonstrated
for Ch.~2 and Ch.~4, respectively.  These levels of uncertainties are
an order of magnitude better than what ground-based weather searches
have accomplished and, even though our targets are significantly
fainter than TrES-1 (and thus the accuracy will be lower), can provide
very constrained limits on the amplitudes of photometric variations in
our targets. We therefore chose to use the staring mode for our brown
dwarf weather program.

The four IRAC channels do not simultaneously view the same position on
the sky, however -- Ch.~1 and 3 view one field of view, and Ch.~2 and
4 view another non-overlapping but approximately adjacent field of
view.  If we are to stare at our target object, therefore, we must
choose which field of view to use.  From an astrophysical point of
view, the choice was not clear cut -- there was no empirical data from
previous IRAC or ground-based observations to suggest that variability
would be greater in one filter-pair, nor was there compelling guidance
from the theoretical models.  We therefore chose to use the same
filter-pair (Ch.~2 and 4) as had been used for the planet transit
observations.  One reason for this is that Ch.~2 is the most sensitive
and the most well behaved (e.g. ``pixel-phase'' effects are thought to
be smaller in Ch.~2 than Ch.~1 -see sec. 5.1-), suggesting that better
relative photometry should be possible with Ch.~2.

If no other constraints were involved, the observations for our
targets would therefore have been extremely simple to describe.  Slew
to the target and center it in the Ch.~2/4 FOV, wait until the
spacecraft pointing has settled, and take $N$ consecutive frames of
data with integration time $M$. Table~\ref{Observations} summarizes
the different settings adopted for each target: observation date,
number of AORs, number of consecutive images taken in each AOR,
integration time per pixel and, total time on target for each observed
object. The integration time is selected so that the number of
electrons in the central pixel is not too large ($\sim$half full well,
so that linearity corrections are small -See the IRAC Data
Handbook\footnotemark); the number of consecutive frames is set so
that the total time on target is significantly greater than the
expected rotation period.  For our first and brightest target,
DENIS0255, however, one additional constraint caused us to deviate
significantly from this simple procedure.  At the time we constructed
the observing plan for DENIS0255, a single AOR was limited to 256
repeated exposures.  Because DENIS0255 is relatively bright, the
maximum exposure time we could use was 12 seconds, hence a single AOR
was limited to of order an hour.  With a desired six hours on target,
we therefore had to break up our observation into six consecutive
AORs.  The only difficulty with this is that the AOR is defined such
that it begins with a slew to the target and a re-acquisition by the
star-tracker, and we could not eliminate that process.  Therefore,
even though for the second through sixth AORs we were already pointed
at the target, there was still a brief acquisition sequence and a
consequent slight repointing of the spacecraft and thus a
repositioning of the target on the arrays at the start of each AOR.

\footnotetext{http://ssc.spitzer.caltech.edu/irac/dh/iracdatahandbook3.0.pdf}

\placetable{Observations}

For 2MA2244, which is much fainter than DENIS0255, the individual
exposure time was instead 100 seconds, and therefore we were able to
observe the target for 6.5 hours with only two AORs. Due to the
background brightness in Ch.~4, the maximum exposure time in this
bandpass is 50 sec.  Hence, we had two 50 seconds exposures in Ch.~4
per each 100~seconds exposure in Ch.~ 2 (104 repeat exposures per AOR
in Ch.~2 and 208 in Ch.~4). For 2MA0908, we were able to avoid these
re-centerings of the spacecraft.  In this case, we conducted the
observations in an engineering mode which had no limit on the number
of repeat exposures, and hence the observation was conducted with
essentially a single AOR and only the initial spacecraft pointing
acquisition.


\section{Data analysis}

Our starting point for the data analysis was the Basic Calibrated Data
(BCD) produced by the IRAC pipeline software (version s13.0.1) at the
Spitzer Science Center (SSC).  The tasks performed by the pipeline are
mainly dark subtraction, multiplexer bleed correction, detector
linearization, flat-fielding, cosmic ray detection and, finally,
flux-calibration. For a detailed description of these processes see
the IRAC Data Handbook\footnotemark[\value{footnote}].  This pipeline
is intended to produce fully flux-calibrated images which have had
most of the well-understood instrumental signatures removed. However
there are some instrumental effects which are not corrected; we take a
close look at them in the next section. The BCD images are calibrated
in units of MJy/sr. Calibrated magnitudes were obtained using the
zero-point fluxes\footnotemark[\value{footnote}] and transforming them
into magnitudes to obtain the appropriate BCD zero-point magnitude for
each channel.  The BCD plate scale used to obtain the zero magnitudes
is 1.22 arcsec/pixel and the zero-points used in the calibration are
listed in Table~\ref{calibration}.

\placetable{calibration}

The finding of a good centroiding and, the photometry extraction were
performed under {\tt IRAF} standard procedures. Both {\tt STARFIND}
and {\tt DAOFIND} routines were used for the source extraction
because, probably due to the pixel-phase effect and the IRAC
undersampling, the routine to derive pixel coordinates within {\tt
  DAOFIND} produced results that were sometimes inaccurate (see next
section).  We performed aperture photometry using {\tt PHOT} with a
source aperture of 4 pixels radius (4.88~arcseconds). The aperture
radius was selected in order to obtain the maximum signal-to-noise
ratio. The sky background was subtracted using an annulus with inner
radius of 15 pixels (18.3~arcseconds) and width of 10 pixels
(12.2~arcseconds). We selected this relatively large sky annuli to
provide the best possible subtraction of background given the lack of
objects close to the targets in our images. The IRAC calibration
aperture has a 10 native pixel radius and thus we had to apply an
aperture correction to our data.  We derived additive aperture
corrections in magnitudes of 0.094 and 0.097 mag for Ch.~2 and Ch.~4
respectively, directly from our own observations.

To compute random errors for our light curves, we assumed that no
significant real variability in our objects occurs on timescales of 20
minutes or less. We measured the scatter of every 10 datapoints (5
datapoints for the faintest object because of the longer exposure
time) and the 1-$\sigma$ error bars in the figures represent the
median of these values. Thus, the errors in the light curves were
computed empirically from the data themselves.  We make no estimate of
the systematic error in our absolute fluxes because our observing mode
is not designed to provide the best absolute fluxes (we are staring at
the target instead of dithering).

\begin{figure}
\centering
\includegraphics[scale=0.35]{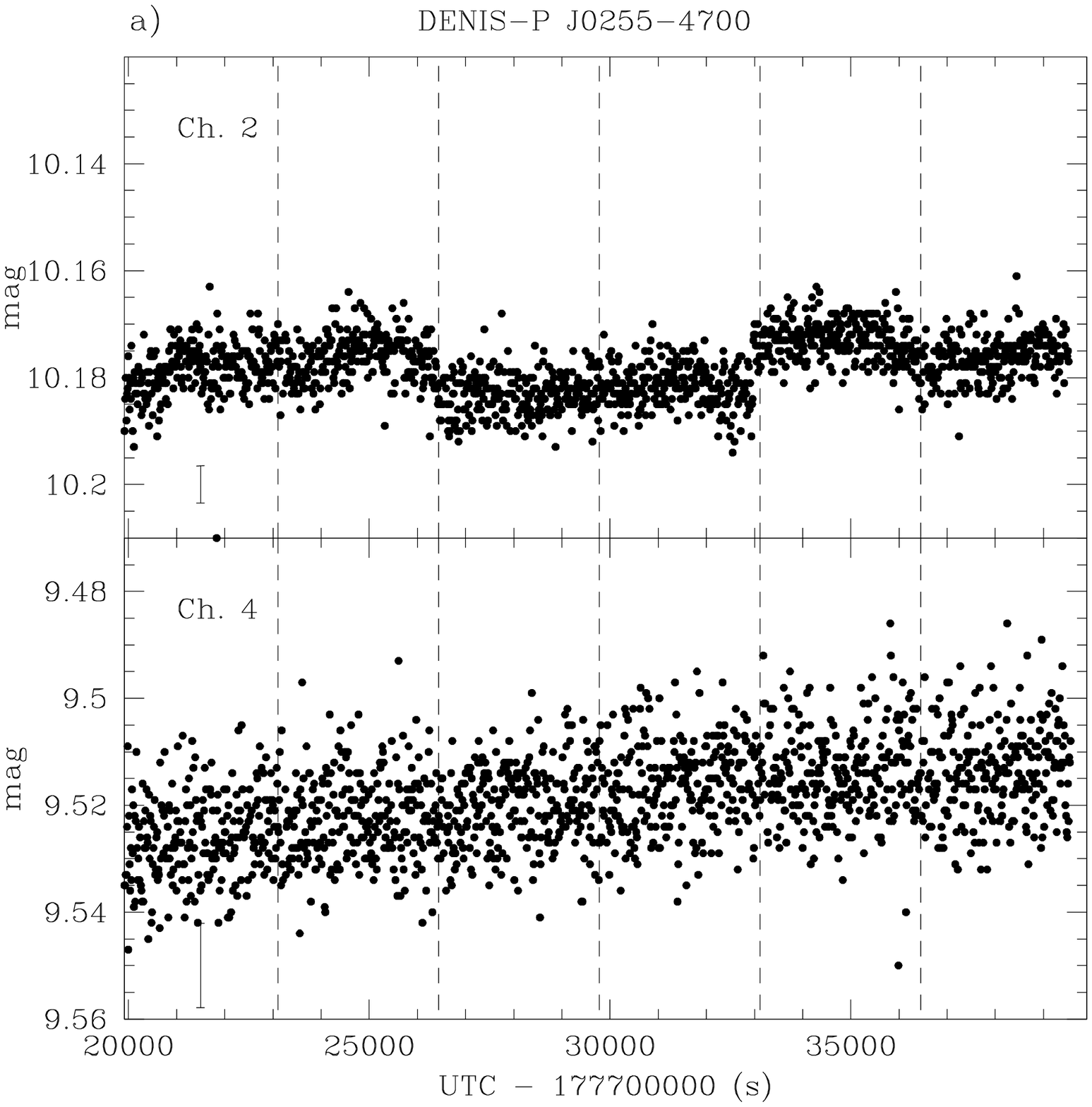}\\
\vfill
\includegraphics[scale=0.35]{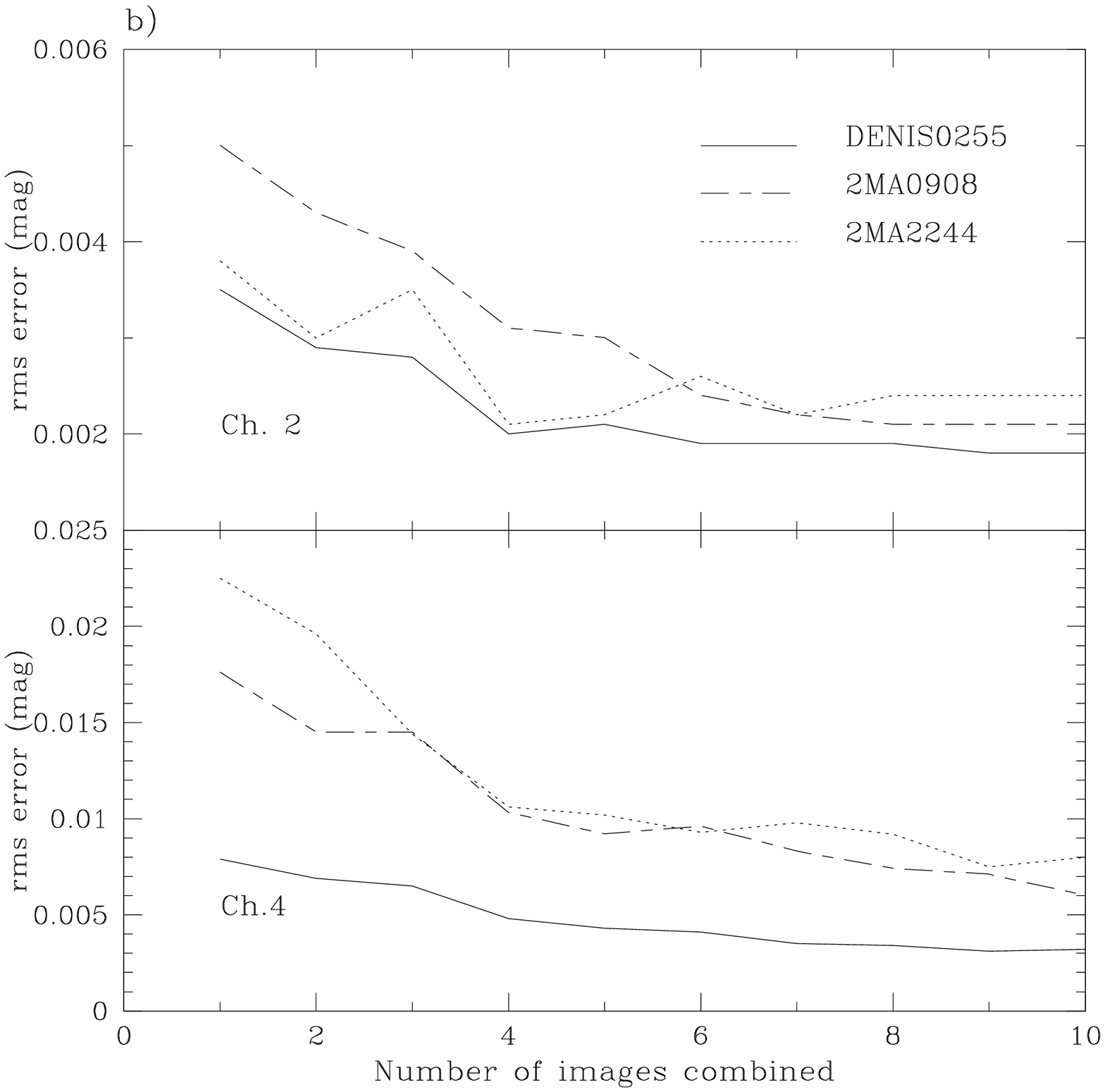}\\
\caption{a) Raw light curve for DENIS0255. The upper and lower panels show the Ch.~2 and Ch.~4 time series respectively and, the vertical dashed lines delimit the different AORs. The one sigma uncertainty per point is represented in the lower left corner of each panel.  b) Dependence of the dispersion about the mean on the number of datapoints combined for DENIS0255 (solid line), 2MA0908 (dashed line) and 2MA2244 (dotted line). Note that the scales are different in the two panels.}
\label{denis_simple_rms}
\end{figure}

\placefigure{denis_simple_rms}

An example of the raw light curves for one of our objects, DENIS0255,
can be seen in Fig.~\ref{denis_simple_rms}a where only the very large,
isolated deviants have been removed ($\sim$ 2 $\%$ of the datapoints,
presumably cosmic ray hits). The Ch.~2 data do show some variation
but, because the changes happen at AOR boundaries, we suspect an
instrumental cause. In order to improve the signal-to-noise, the BCD
images were combined in groups. We selected 5 as the number of images
to combine in each group for our final analysis as a trade off between
maximizing the signal-to-noise ratio of source flux (see
Fig.~\ref{denis_simple_rms}b) while at the same time preserving
temporal resolution. Therefore, we have 300 merged datapoints with 1
minute increments spanning almost six hours of observation time for
the first target, DENIS0255. For 2MA0908, the observations were taken
under only one AOR spanning approximately 8 hours. After combining the
images, we had 178 datapoints in increments of 2.5 minutes. Finally,
we have 42 datapoints in 8.3 minute increments for the faintest
target, 2MA2244, which was observed for 6.5 hours in two different
AORs.  Because we have double number of images in Ch.~4 (half exposure
time each) than in Ch.~2 (see Sec.~3), we combined the images in
groups of ten for the Ch.~4 data to match the time increment in both
channels.  We have at least one field object per target and they were
analyzed in exactly the same way as the science targets. However, we
did not perform differential photometry because, even though the 2MASS
$K_s$ magnitudes of the field objects were comparable to those of our
targets, their IRAC magnitudes were significantly fainter (between 1
and 3 mag fainter) and therefore their light curves were much noisier.
We did use them as control objects, comparing their time series with
the science ones.

\begin{figure}
\centering
\includegraphics[scale=0.25]{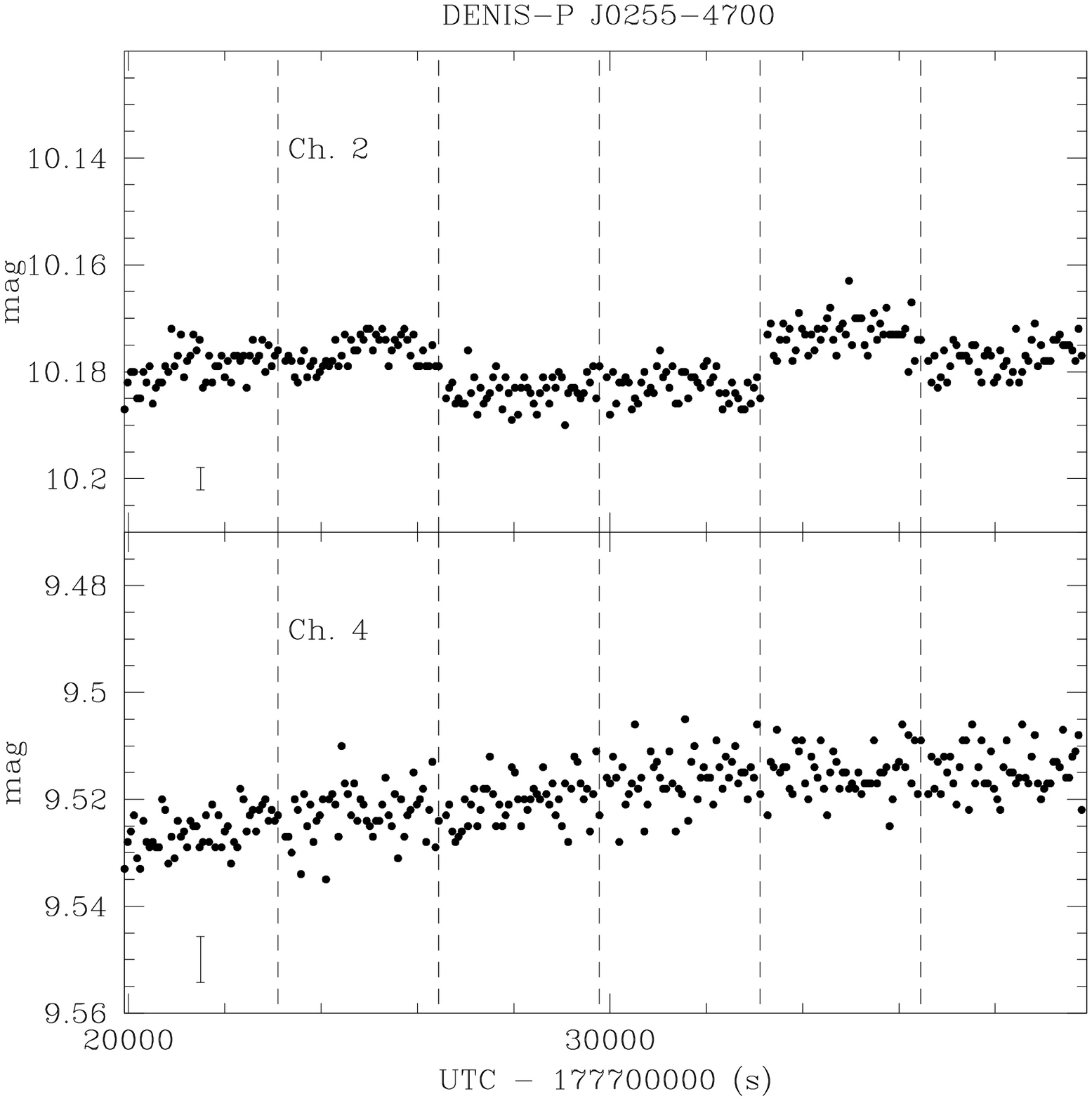}\\
\vfill
\includegraphics[scale=0.25]{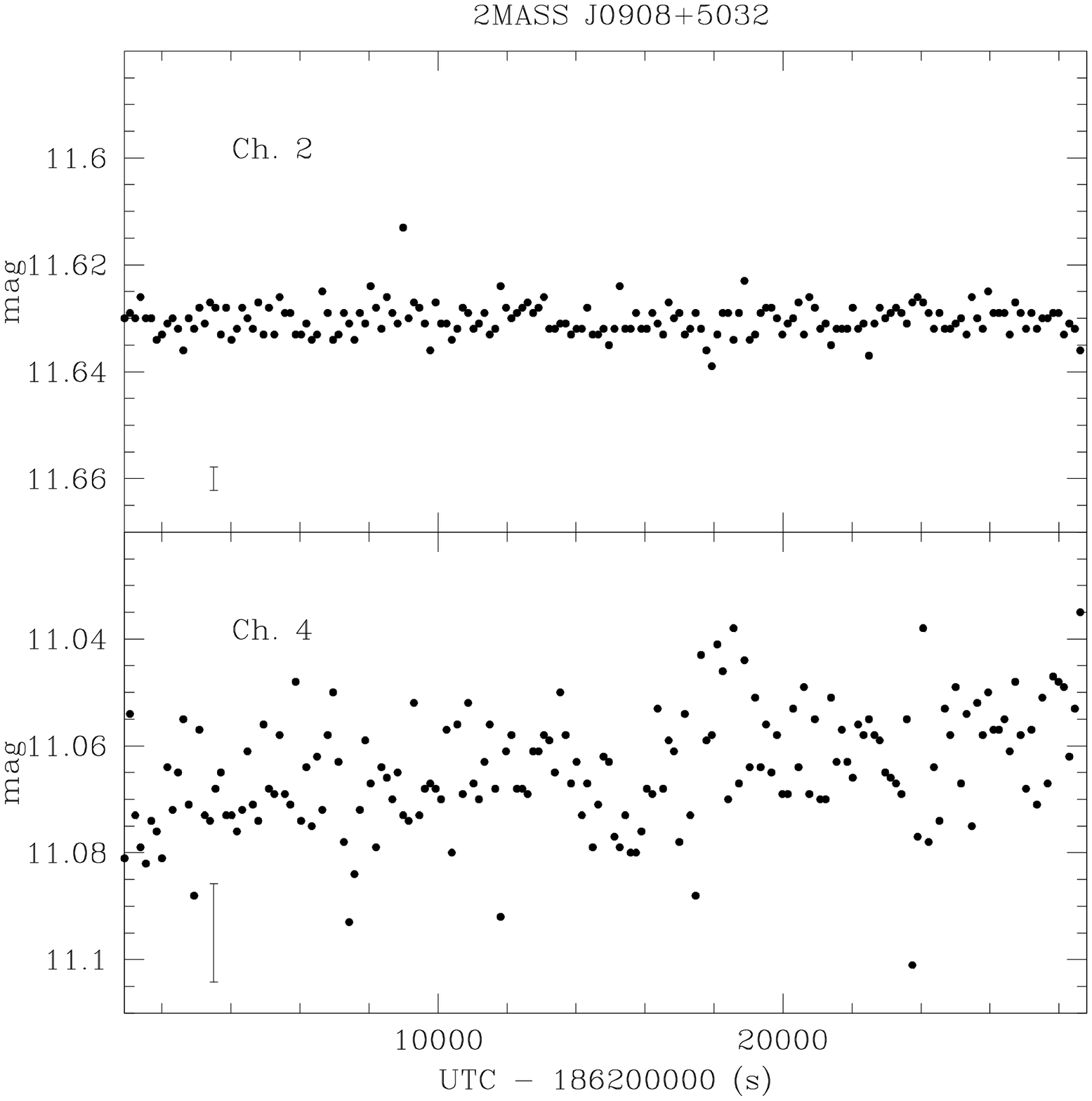}\\
\vfill
\includegraphics[scale=0.25]{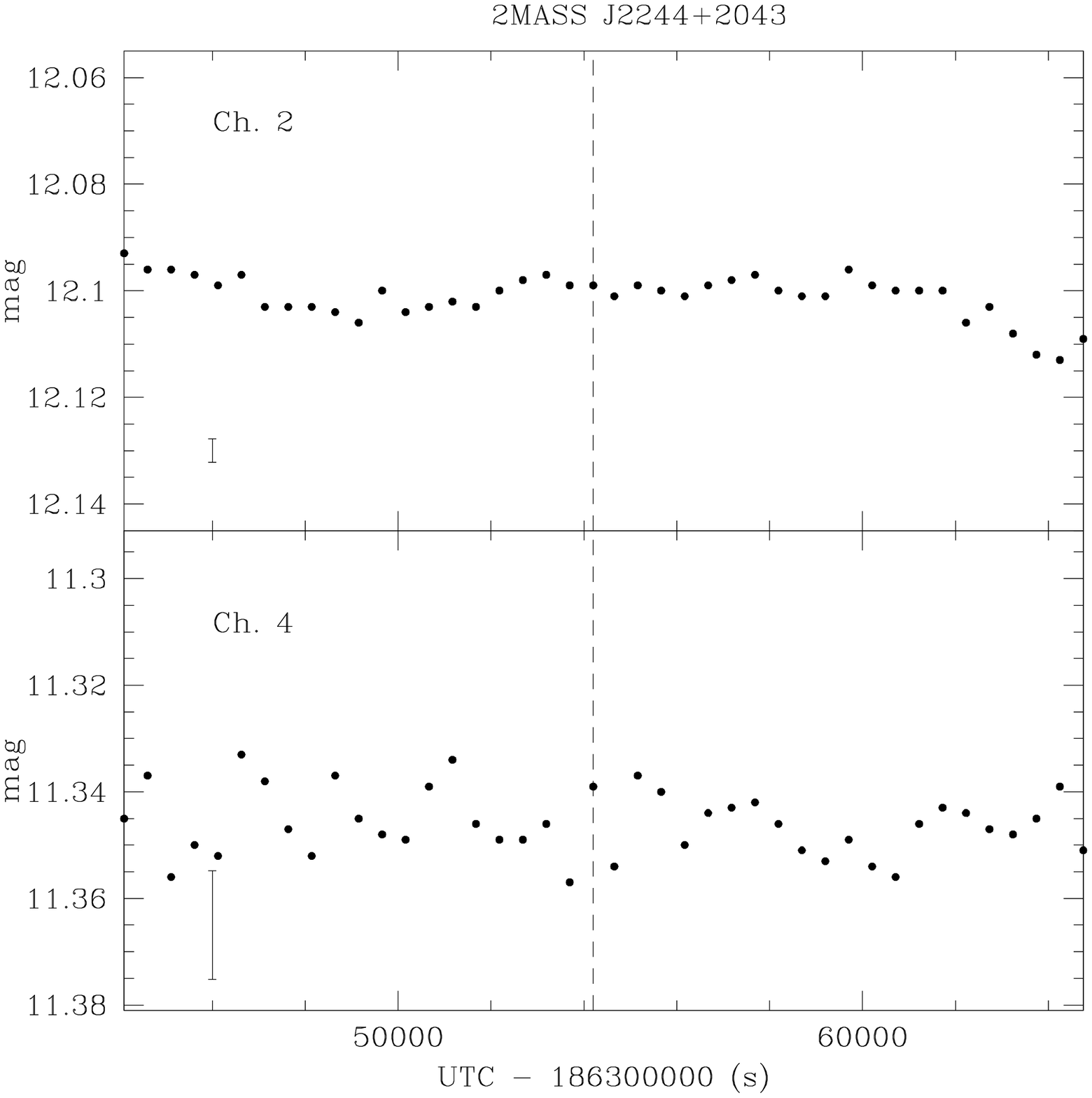}\\
\caption{Light curves for DENIS0255, 2MA0908 and 2MA2244 obtained
  from the binned data. The upper and lower panels show Ch.~2 and
  Ch.~4 time series respectively. The RMS-uncertainty of a single
  binned point is represented in the lower left corner of each panel.
  The vertical dashed lines delimit the different AORs.}
\label{lc_raw}
\end{figure}

The time series for the averaged datapoints for our three targets in
both bandpasses can be seen in Fig.~\ref{lc_raw}. The upper panels are
the light curves for Ch.~2, extracted as explained above,
and the lower ones are for the Ch.~4 data. In these time
series, without any possible corrections applied, we see no evidence
of a rotational variability (at least in DENIS0255 and 2MA0908). Upper
limits on the intrinsic variability of our targets at Ch.~2
and Ch.~4 bandpasses were established as the RMS of the light
curves. Therefore, if any sinusoidal variation is present its
RMS amplitude would be below 5, 3 and 4 mmag for DENIS0255, 2MA0908
and 2MA2244 respectively in Ch.~2, and below 6, 10 and 6 mmag in
Ch.~4.

\placefigure{lc_raw}

\begin{figure}
\centering
\plotone{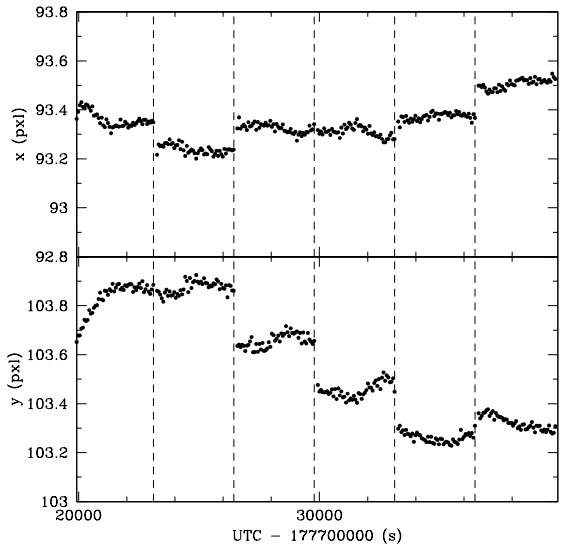}
\caption{{\tt STARFIND} centroid position as a function of time for DENIS0255. The $x$ and $y$ positions are shown at the top and the bottom panels respectively. The vertical dashed lines denote the boundaries between AORs. The spacecraft re-acquires guide stars at the start of an AOR, causing the disjointed jumps in position.}
\label{denis_centroid}
\end{figure}

The Ch.~2 data do show photometric variations, particularly in
DENIS0255, as illustrated in the upper left panel of
Fig.~\ref{lc_raw}.  Those variations are clearly correlated with the
change with time of the star's centroid position (see
Fig.~\ref{denis_centroid}), and are the most noticeable in DENIS0255,
with a maximum amplitude of 1-2\%.  The light curve of this object
exhibits some large discontinuities that occur at the transition from
one AOR to the next.  The correlation between centroid position and
photometric variations is not so obvious for the two fainter objects,
but this is probably due to the fact that their movement is much
smaller, $<$0.1 pixels for 2MA0908 and around 0.3 pixels for 2MA2244.
Looking at the centroid position versus time, the pointing jitter is
very small inside a single AOR, but offsets as large as 0.7 pixels
occurred along the whole observation period due to the re-acquisition
of guide stars at the beginning of each observation.  In the case
DENIS0255, we also found a large drift, about 0.2 pixels (0.24 arcsec)
in the target's $y$ position on the array during the first 20 minutes
of the first AOR. (See Sec.~5.4 for a further discussion of the
pointing variations and their influence on the Ch.~2 photometry.)

\placefigure{denis_centroid}

The first step in deriving time series photometry is the determination
of the centroid positions for the target star in each image.  We
initially used {\tt DAOFIND} for this purpose, but noticed odd shifts
(large shifts and even bimodal positions) in the centroids for some
images which we believed to be spurious.  We wrote a simple
first-moment routine to check the {\tt DAOFIND} centroids, which
worked better with the centroiding but was relatively noisy.  We
finally settled on the {\tt STARFIND} routine, which we believe
returns good centroids for nearly all of the images.  The
undersampling in the Ch.~2 makes the centroid determinations
relatively inaccurate even for {\tt STARFIND}, but we do not believe
the trends in the light curves are a result of this imprecision.  If
the photometric variations were primarily attributable to errors in
the centroiding, increasing the aperture size would have helped.
However, we found a similar trend using bigger apertures, with the
only difference being, noisier light curves depending on the aperture
we used. Moving the sky annulus further out did not remove the effect
either.

The discontinuities in the Ch.~2 photometry at AOR boundaries
could also be due to pixel-to-pixel flat-field errors in combination
with the position shifts illustrated in Figure~\ref{denis_centroid}.
We examined the flat-field used, and there are differences in the
values of the flat-field of order 2\% between different pixels near
the location of DENIS0255 which could, in principle, cause the
photometric shifts we see in Figure~\ref{lc_raw}.  As a test of this, we
extracted photometry from the raw data frames and found a light curve
very similar to that derived from the BCD data.  This does not
completely exclude flat-field errors as the cause of the variations
seen for DENIS0255 in Figure~\ref{lc_raw}, but we believe this is not a
significant contributor.

The Ch.~4 data do not show the same photometric variations as the
Ch.~2 data.  Instead, DENIS0255 and 2MA0908, the two brightest
objects, show brightening of 1.5\% along the whole observation period.
We discuss this effect in Sec.~5.2.


\section{Instrumental effects and corrections}

\subsection{Pixel-phase effect}

The number of electrons created in the image of a star in IRAC Ch.~1
and Ch.~2 depends on exactly how the star is centered relative to the
center of a pixel.  This effect is probably the result of light losses
at the boundaries between pixels. It is repeatable, and there is a
quasi-linear relation between what we measure as the magnitude and the
displacement from the center of the pixel.  Therefore, a star whose
image is centered on the center of a pixel has the maximum apparent
flux, while a star centered on the interstices of four pixels has the
minimum apparent flux. This effect is called the ``pixel phase
effect'' and more information is available in the IRAC Data Handbook.

This artifact results in a variation in the detected flux of an object
as its image moves relative to the center of a pixel.  The lack of a
detectable pixel phase effect for IRAC Ch.~3 and Ch.~4 is probably
due to their use of a different detector technology (SiAs vs.\ InSb)
and to the broader PSFs for the longer wavelength channels.

The SSC provides a functional form for the correction for pixel phase
effect for Ch.~1 on its website.  Pixel phase is defined as the
distance of the centroid position of a star from the center of the
pixel with the most flux, thus:
\begin{equation} 
phase = \sqrt{(x-x_o)^2 + (y-y_o)^2}
\end{equation}
where, for each image, {\it x, y} are the positions of the source's
centroid and \begin{math} x_o, y_o\end{math} are the integer pixel
numbers containing the source centroid.  The correction for Ch.~1 is
defined as a linear relation in flux:
\begin{equation}
Correction=1+0.0535\times\left[\frac{1}{\sqrt{2\pi}}-phase\right]
\end{equation}

The SSC does not provide a similar formula for Ch.~2 because the
scatter in the data available to calibrate the effect is comparable to
the effect.  The FEPS (Formation and Evolution of Planetary Systems)
legacy team has also examined their IRAC BCD images for $>$ 300 nearby
F-, G-, and K-type dwarfs for pixel phase effects.  They find a very
similar relation for Ch.~1 as the one provided by the SSC.  For Ch.~2,
they also find ambiguous data.  For some positions on the array, they
see a similar pixel phase relation as for Ch.~1; at other positions,
they see no obvious pixel phase effect (Meyer, Hillenbrand, Backman et
al. 2004).

\begin{figure}
\centering
\plotone{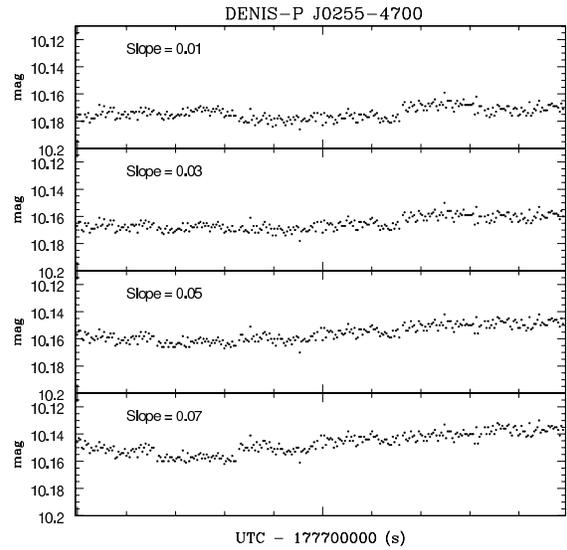}
\caption{Different pixel-phase corrected light curves for DENIS0255 
  depending on the slope adopted in the pixel-phase correction.
  We adopted a slope of 0.05 as the best one.}
\label{slope}
\end{figure}

We chose to assume that a pixel-phase effect might be present in our
Ch.~2 data, and to determine empirically the size of the effect (the
slope of the relation between pixel-phase and flux).  We modeled the
effect as a linear relation between flux and pixel phase, varied the
slope of the relation from 0.00 to 0.07, and examined the light curves
for our three L dwarfs and the field objects for each choice of slope.
We assumed that the slope that minimized the discontinuities in the
photometry between the AORs was correct.  This led us to a slope of
0.05 -- very similar to what is found in Ch.~1.  Figure~\ref{slope}
shows the light curves for DENIS0255 for several different choices of
the slope to the pixel phase correction formula.

\placefigure{slope}

\subsection{Latent image charge buildup}

The two brightest objects of our sample show an upward trend in
brightness of 1.5\% from the beginning to the end of the
observation at Ch.~4 (see Fig.~\ref{lc_raw}). The shape of the
light curves is very different from what we see in Ch.~2 and, if
real and interpreted as rotational modulation, would imply periods much
longer than those inferred from the spectroscopic rotational velocities.  We
believe instead that what we are seeing in the Ch.~4 data is a
latent image buildup.  This effect was also observed in
\citet{Charbonneau05}, where the target and calibrators were brighter
than our targets.

This instrumental effect may depend on the flux of the target. In
addition, there is a pixel dependent term in the behavior of the long
term latents, and it is possible that they are frametime
dependent. Despite that, and even though the non-variable calibrators
in \citet{Charbonneau05} data are brighter than our targets, there is
no other dataset more similar to ours in terms of time staring to an
object, so we decided to use their calibrators to correct the
photometry of our targets in Ch.~4. We reanalyzed their BCDs,
extracted the photometry, and used the normalized flux to fit a second
degree polynomial to each calibrator.  Then the time series of
DENIS0255 and 2MA0908 were divided by the mean of both fittings.

The functional form for this correction is: 
\begin{equation} 
Correction=-2.2402\cdot10^{-11}\times t^2+1.1872\cdot10^{-6}\times t+0.9917
\end{equation} 
\begin{equation} 
Corrected~Flux (MJy/sr)=Flux/Correction
\end{equation} 
where $t$ is the time (in seconds) when the exposure was taken
assuming the first exposure occurred at $t=0$.

Our faintest object -- 2MA2244 -- does not show an increase in its
brightness with time for the Ch.~4 photometry and thus, we did not
apply the correction to this object.  The difference in the latent
behavior in this case is probably due to the different frametimes used
for this object and the fact that two repeats of 50~second each are
used to synthesize a 100~second frame. Different frametimes have
slightly different commanding that leads to small differences in the
delay between consecutive integrations.  It is possible that 2MA2244
does not show a significant latent buildup because the latent images
are sensitive to such delays. The dependence of latent charge buildup
as a function of position on the array, frametime and flux would have
to be studied before a more accurate correction could be applied.

\subsection{Periodic movement of the pointing}

\begin{figure}
\centering
\plotone{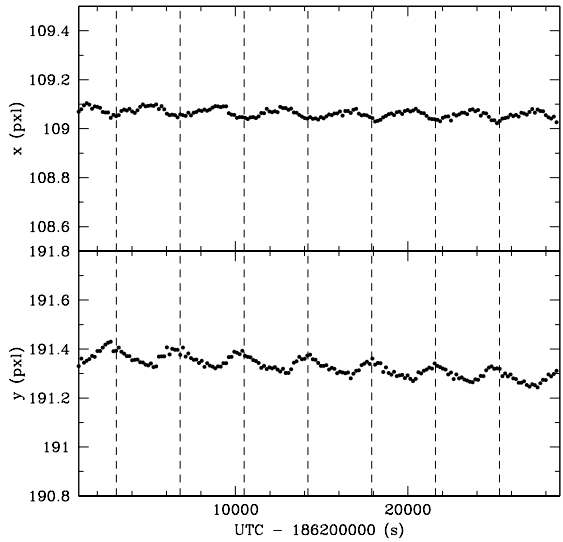}
\caption{Array $x$ position (upper panel) and $y$ position (lower panel) as a function of time for 2MA0908. Both positions oscillate with a period of $\sim$ 3000sec. Small heaters near the star tracker cycle their power with a similar period and are likely producing flexure in the trackers.}
\label{saw-tooth}
\end{figure}

Since the observation of 2MA0908 was performed under only one AOR, it
gave us the opportunity to study the pointing without the large shifts
introduced by the change of AORs. In this case, the movement of the
$x$ and $y$ positions with time showed a saw-tooth pattern with a
period of 3000~sec and a peak to peak amplitude of 0.1 pixel (see
Fig.~\ref{saw-tooth}), with the largest amplitude in the $y$-axis of
the array. There is also a slow, approximately linear drift in the
$y$-axis position, amounting to approximately 0.1 pixel over the 8
hour period of the observation.

We examined the pointing history file for the time period while our
targets were being observed and there was no measurable telescope
oscillation.  There is a small, approximately constant drift in RA
during the observation, and a small pointing discontinuity when a new
AOR starts, but we do not see the 3000 sec period that we see with the
IRAC data. There are temperature sensors attached to the cold plate on
which IRAC is mounted. The sensors indicate an oscillation in
temperature with a similar period. The heaters located near the star
tracker could be cycling on and off, causing the tracker to bend
slightly, and that could be a plausible cause for this effect.

In any case, the effect of this oscillation on the light curves is
very small and it should be fixed with the pixel phase correction
applied.

\placefigure{saw-tooth}

\subsection{The Corrected Photometry}

\begin{figure}[tbp]
\centering
\includegraphics[scale=0.25]{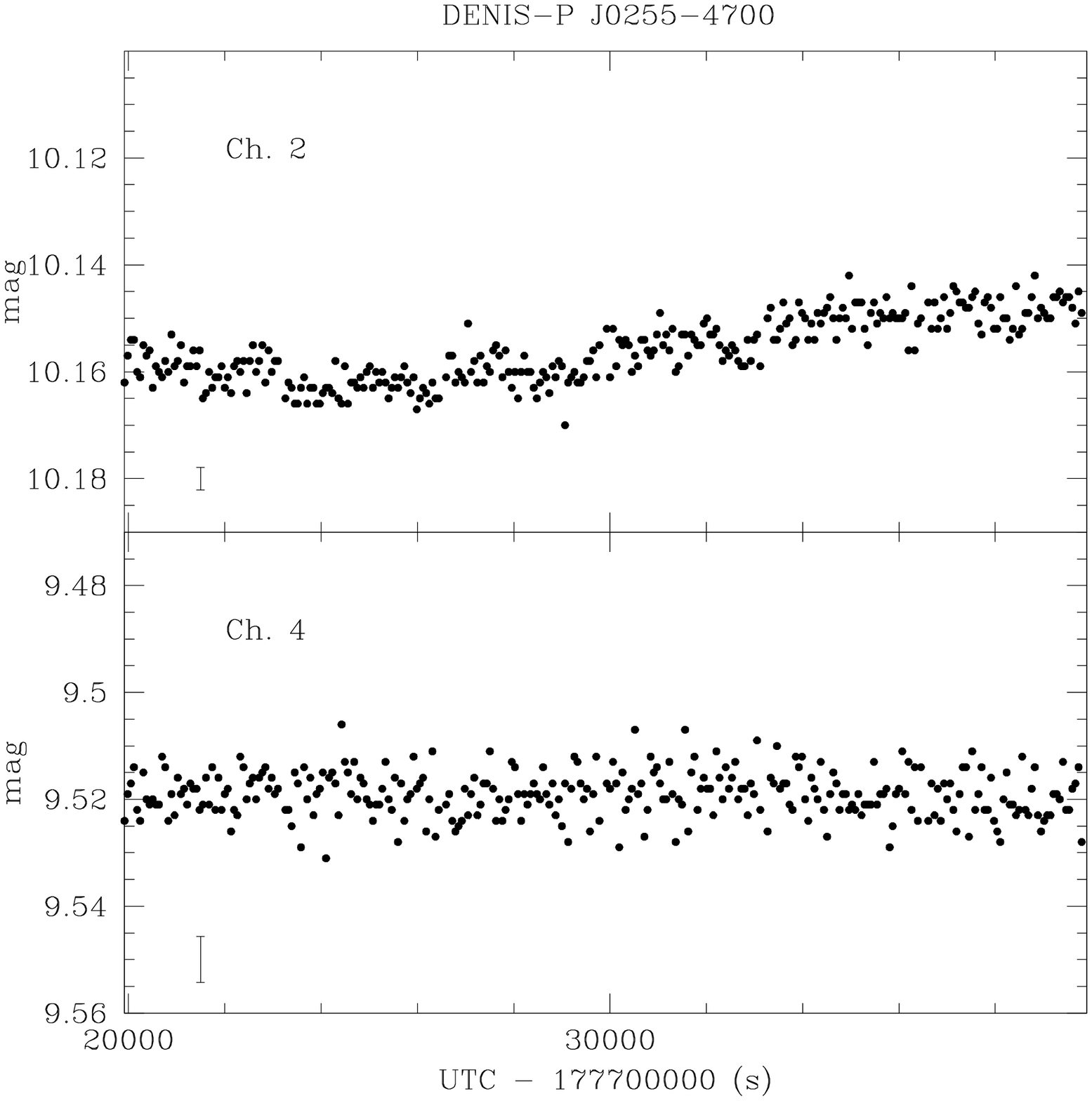}\\
\vfill
\includegraphics[scale=0.25]{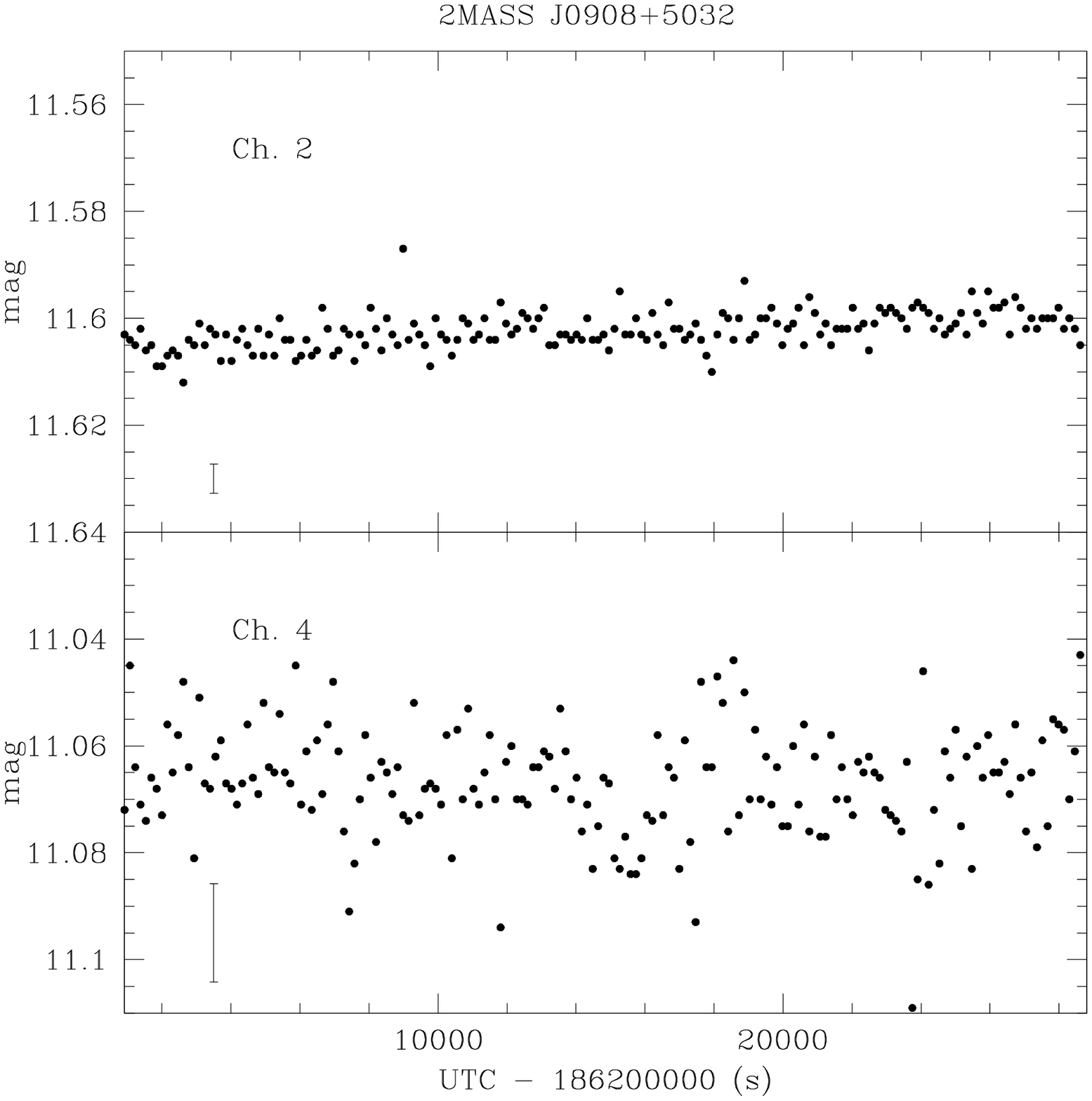}\\
\vfill
\includegraphics[scale=0.25]{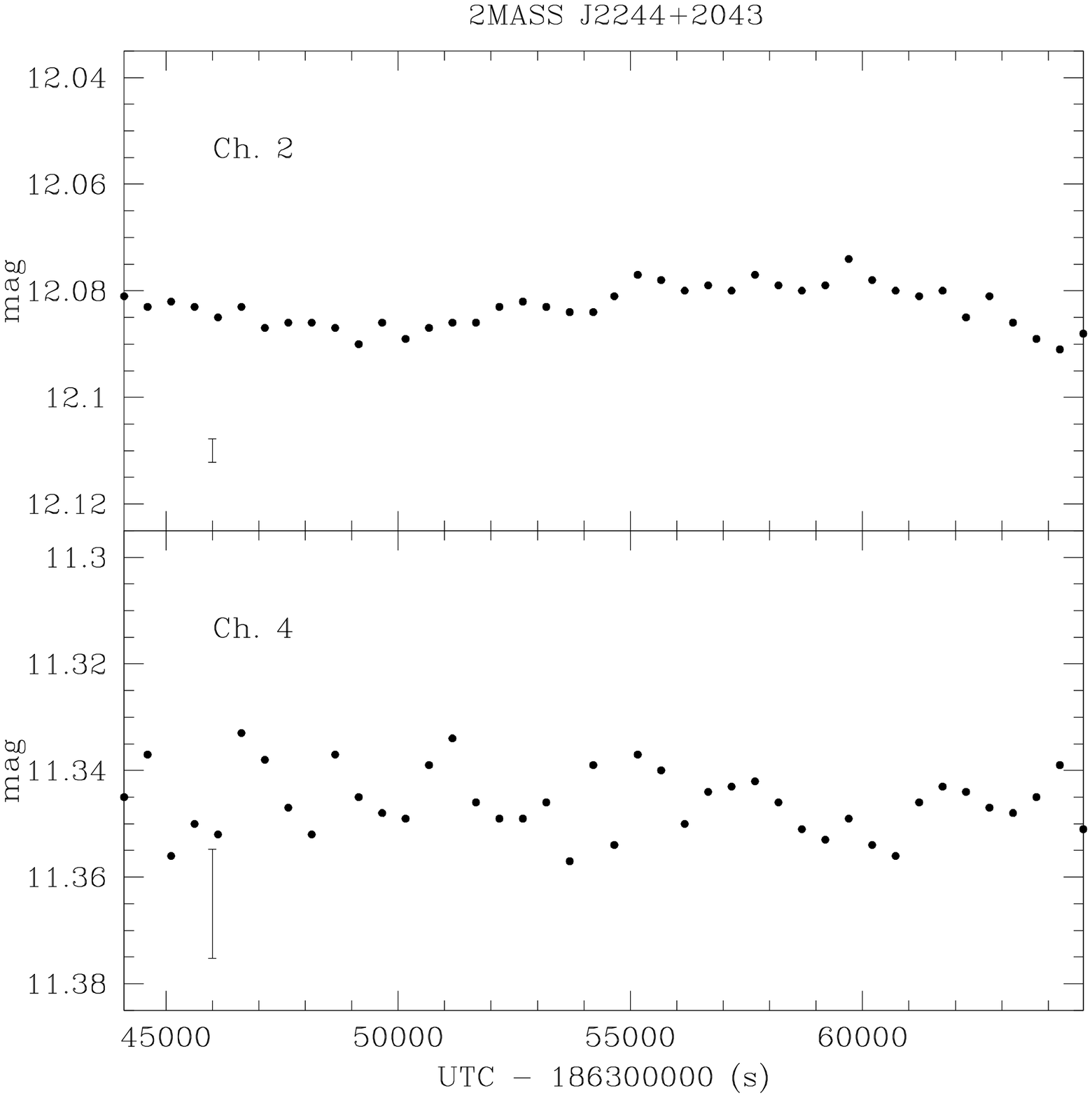}\\
\caption{Final light curves for DENIS0255, 2MA0908 and 2MA2244. 
The three of them have been corrected from pixel-phase at 
Ch.~2 (upper panels) and DENIS0255 and 2MA0908 have 
been corrected from latent images at Ch.~4 (lower panels).
The one sigma per point uncertainty is represented in the 
lower left corner of each panel.}
\label{lc_corr}
\end{figure}

Figure~\ref{lc_corr} shows the light curves of the three targets,
corrected for the effects of pixel phase and latent image buildup. The
upper and lower panels show the Ch.~2 and Ch.~4 data respectively. The
RMS-error is represented by an error bar at the lower left corner of
each panel.  After applying the pixel-phase correction to Ch.~2 data,
discontinuities between AORs are no longer visible. The photometry of
two brightest objects, DENIS0255 and 2MA0908, were corrected for
latent images in Ch.~4 (2MA2244 did not show that effect probably due
to its faintness and different frametime) and now appear flat.  Note
that the trends in both bandpasses are different and that, at least
for DENIS0255 and 2MA0908, there is a lack of photometric modulation
at the expected rotational periods. The RMS of the light curves are 6
and 4 mmag for Ch.~2 and Ch.~4 respectively for DENIS0255, 3 and 9
mmag for 2M0908 and, 4 and 8 mmag for the faintest object, 2MASS2244.
Therefore, any possible variability on the timescale of 6 or 8 hours
would be less than these values.

\placefigure{lc_corr}


\section{Analysis of variability}

The data of each brown dwarf was analyzed in a similar way to that of
\citet{Bailer99}. The $\chi^2$ test was used to determine the
probability that the deviations in the light curve are consistent
with the photometric errors (i.e. non-variable). The null hypothesis
for the test is that there is no variability. We evaluated the
$\chi^2$ statistic:
\begin{equation}
\chi^2=\sum_{k=1}^{k=K} \left(\frac{\Delta m(k)}{\sigma}\right)^2
\end{equation}
where $K$ is the number of datapoints in the light curve, $\Delta
m(k)$ is the magnitude for each datapoint with the mean magnitude
subtracted and, $\sigma$ is the RMS-error.

A large $\chi^2$ value indicates a greater deviation compared to the errors
and thus, a smaller probability that the null hypothesis is true (i.e.,
variable). This probability, $p$, is calculated and we will claim
evidence for variability if $p < 0.01$ (a $2.5~\sigma$ detection).
This method is very sensitive to the accuracy of the errors. We
believe that the technique used to estimate the errors (obtained
empirically from the data themselves) has the advantage that false
detections associated with underestimating the errors can be avoided.

If evidence of variability was found in an object, we looked for a
periodic signal in the data following the methodology described by
\citet{Scargle82}.  This method is equivalent to a least-squares fit
(in the time domain) of sinusoids to the data. The algorithm
calculates the normalized Lomb periodogram for the data and gives us a
false-alarm probability based on the peak height in the periodogram as
a measure of significance.

We also examined carefully the data in order to identify any possible
signal that could be interpreted as the result of a brown dwarf flare.
However, only single-point (before binning) deviants -presumably
radiation events in the detector- were found.

\section{Results and discussion}

\subsection{DENIS-P J0255-4700}
DENIS-P J0255-4700 is the brightest member of our sample.  That and
its late-L spectral type make this target perfect for this study.
Furthermore, it is one of the best studied objects in the
late-L/early-T region. It has been claimed to be variable in the {\it
Ic} band on more than one timescale \citep{Koen05} but, on the other
hand, no signs of variability have been found in any other band.  This
object has a $v\sin{i}$ of $\sim$40~km~s$^{-1}$
\citep{Basri00,Zapatero06} and hence, its rotation period should be of
3~hours or less and our 6~hours of continuous observation should
capture two full periods.

Table~\ref{results} shows main results for all targets including IRAC
magnitudes, RMS-amplitudes, probability of an object to be
non-variable and, period of the modulation observed. This object was
labeled as variable in Ch.~2 ($p \le 10^{-4}$) and, non-variable in
Ch.~4 ($p = 0.3$) by the criteria used. However, if any variability is
present, it has to be under a RMS-amplitude of 6 mmag for Ch.~2 and 4
mmag for Ch.~4 (See top panel of Fig.~\ref{lc_corr}).  The periodogram
searches for periods in the interval ranging from that corresponding
to the the Nyquist frequency ($\sim$3~min) to values slightly larger
than the interval covered by our observations. The power spectrum of
this object shows only one strong peak at ~7.4~hr, almost twice the
period predicted from the spectroscopic rotational velocity.  Hence,
the cause of variability would have to be some type of global change
in the luminosity of the object (which for some reason is not
modulated on the rotation period). Future observations would be useful
in order to determine if any kind of long term variability is present.
Another possibility would be that the $v\sin{i}$ is in error or that
our assumed radius is in error (in both cases by of order a factor of
two). However, recently \citet{Zapatero06} have derived the same
$v\sin{i}$ with higher accuracy by using Keck/NIRSPEC IR
spectrograph. DENIS0255 doesn't show any evidence of lower gravity in
its optical spectrum or near-IR colors and thus, nothing indicates
that it has a larger than normal radius (as might be the case if it
were very young). Note that our 6~hr of observation do not allow us to
see an entire phase and thus, we cannot check the validity of the
estimated period.

\placetable{results}

We note that our DENIS0255 observations had by far the largest
movement in the stellar centroid during the observing period of our
three targets.  We know that there are instrumental effects that
depend on position on the array (both pixel phase effects and
flat-field errors) that affect the measured flux in Ch.~2, and those
effects are smaller for Ch.~4.  Therefore, even having removed the
instrumental effects, the most likely object for us to see a spurious
signal for was DENIS0255, and we should have seen it to be larger in
Ch.~2 -- exactly as was the case.

On the other hand, the fact that we see variations in Ch.~2 and
not in Ch.~4 is not inconsistent with the hypothesis of real
variability arising from clouds.  The spectra of L and T dwarfs are
sculpted by molecular absorption bands which vary greatly in strength
as a function of wavelength.  Thus, there is no well defined
``photosphere,'' and the depth from which flux is emitted varies
strongly with wavelength.  Assuming a well-defined cloud layer, flux
may originate from above, within, or even (for small optical depths)
from below the cloud layer \citep{Ackerman01}.  Thus if a local hole
suddenly appears in an otherwise uniform, global cloud deck, it will
only be apparent at those wavelengths that would otherwise originate
from within or below the cloud.  The presence of the hole would not be
apparent in spectral regions originating from well above the cloud
deck.  This effect is well known from observations of Jupiter.  The
``five micron hot spots'' \citep*{Westphal74} of Jupiter arise from
holes in the global ammonia cloud deck, allowing flux from hotter,
deeper-seated regions to escape to space.  The hot spots are apparent
at 5~$\mu$m because this is a region of relatively low molecular
opacity.  These hot spots are not apparent at longer wavelengths where
flux originates from higher in the atmosphere.

Among the IRAC bandpasses, Ch.~1~\&~2 (3.6~\&~4.5~$\mu$m) probe most
deeply into late L dwarf atmospheres.  Because they overlap regions of
higher molecular (primarily water and carbon monoxide) opacity,
Ch.~3~\&~4 (5.8~\&~8~$\mu$m) probe higher in the atmosphere, generally
above the region cloud models predict is occupied by the iron and
silicate clouds \citep{Ackerman01, Marleyprep}. All else being equal,
we expect any variability arising from non-uniform cloud coverage to
be greatest in Ch.~1~\&~2.  If the dispersions observed in the Ch.~2
data do in fact arise from atmospheric variability, we predict that
comparable or larger variations would be detectable in Ch.~1, but not
Ch.~3.

Whether the 7.4 hour modulation in Ch.~2 is instrumental in origin or
intrinsic to the target, our data place a limit on the amplitude for a
true rotational modulation with a period between 20~minutes and
6~hours below 6~mmag for this channel.

\subsection{2MASS J0908+5032} 

This object has very discrepant optical and near-infrared spectral
types that could indicate a cloudy atmosphere.  Its $v\sin{i}$ is
31~km~s$^{-1}$, thus its period should be less than 4 hours and our
observation would again obtain two whole periods.

A glance at the light curve of 2MASS0908 should be enough to convince
the reader that coherent rotational modulation is not present. This
object shows no prominent features in its light curve more than a very
slight increment of its brightness along the whole observation period
for Ch.~2.  Again this pattern is not confirmed by the Ch.~4 data and
thus, it seems that some other cause, aside from intrinsic
variability, is responsible for the feature.  The $\chi^2$ test labels
this object as non-variable in both channels. Any possible variability
over the 8 hours is at or below the 3 mmag and 9 mmag level in Ch.~2
and Ch.~4, respectively.

\subsection{2MASS J2244+2043}

2MASS J2244+2043 is a L7.5 brown dwarf, with very red near-infrared
colors that could be indicative of dust in its atmosphere. We do not
have a measure of the $v\sin{i}$, but based on the mean $v\sin{i}$ for
L dwarfs, we expect a rotational period of approximately 6.5~hours or
less. 2MA2244 is the faintest object in our sample.

The results of the $\chi^2$ test indicate variability in Ch.~2 and no
variability in Ch.~4. Again, as in DENIS0255, the Ch.~4 data does not
show the same trend.  Indeed, its light curve at Ch.~2 (last panel in
Fig.~\ref{lc_corr}) shows a small amplitude, approximately sinusoidal
modulation. If the variation is intrinsic to the target, a feature in
the brown dwarf's atmosphere, or some differences in the cloud
covering fraction could be causing it.  However, such differences should
be very small since the RMS-amplitude of the light curve is only 4
mmag.  The periodogram of this object shows again only one strong peak
at 4.6~$hr$. This value is consistent with the range of rotation
periods expected for this object. However, note that even though the
variations of the centroid for this object are much smaller than those
of DENIS0255, there was still a bump of 0.3~pixels in the transition
of AORs (just in the middle of the observation period).

\subsection{Limits on Non-Axisymmetric Cloud Distributions for our Targets}

Atmospheric clouds (or other surface inhomogeneities) affect the
observed photometry due to the lower luminosity of the cloud in
comparison with that of a free-cloud region of equivalent size. We
have made a simple model to constrain the size of the feature that
could be causing the observed variability (other models have been
presented by \citet*{Clarke03,Bailer02}). The proposed scenario is an
L8 brown dwarf with a small inclination to the line of sight and a
spot or group of spots at low latitude in its atmosphere. We assume
that, if the cloud deck starts to break up, the cloud-free parts would
have spectral characteristics like those of an early T dwarf. Assuming
typical $J-[Ch.~2]$ colors for both kind of objects we can derive the
difference in brightness and hence, the approximate size of the spot
that could be causing the observed amplitude.  From DENIS0255 data, we
can say that the maximum photometric amplitude of a half sine-wave
light curve would be 6~mmag in Ch.~2 and hence, at this level of
approximation, we can place a rough limit of a spot size of $\sim$1\%
of the visible hemisphere of the object. The same calculation for
2MA2244 leads to a limit of a spot size of $\sim$2\% of the visible
hemisphere.


\section{Summary and conclusions}

We have conducted a photometric monitoring program of 3 late-L brown
dwarfs at the Ch.~2 (4.5~$\mu$m) and Ch.~4 (8~$\mu$m) bandpasses with
observations that lasted for one or two rotational periods of the
object. This project presents the most sensitive search yet obtained
for brown dwarf mid-IR variability. The observational mode selected
allowed us to obtain very well-sampled light curves in the time domain
and 1$\sigma$ RMS uncertainties of $<$3~mmag in Ch.~2 and around
9~mmag in Ch.~4. For each target brown dwarf, the search was sensitive
to the timescale of our observations (6 or 8 hours depending on the
object) and hence, larger variability on timescales to which we were
not sensitive could be present.

Two out of the three objects studied exhibit some variation in their
light curves. DENIS0255 turned out to be variable in Ch.~2 according
to the $\chi^2$ test, with a 99$\%$ confidence level. A period of
7.4~hr was derived using the normalized Lomb periodogram. If this
variability is real and if it is a rotational modulation, its period
would be much larger than the rotational period and would have a
peak-to-peak amplitude of 10~mmag. The cause of variability could also
be some type of global longer term change in the luminosity of the
object which for some reason is not modulated on the rotation period.
The fact that some instrumental effects that could affect the
photometry at Ch.~2 were larger in DENIS0255 than in any other object
suggests that perhaps the variability is not real. The Ch.~4 data show
a flat light curve with no possible variability over the 4~mmag level.
However, since the flux at the two bandpasses arises from different
vertical regions in the atmosphere, the different shapes in the light
curves for Ch.~2 and Ch.~4 are consistent with the hypothesis of
variability caused by clouds in the atmosphere of the L dwarf. 2MA2244
was also labeled as variable by $\chi^2$ test. In this case, its
derived period of 4.6~hr is compatible with the expected rotational
period. This photometric modulation would have a peak-to-peak
amplitude of 8~mmag. Note that the expected period for this object
comes from a mean $v\sin{i}$ for L dwarfs and thus we cannot prove
that there is a rotational modulation with these data.  Again the
feature is not confirmed by the Ch.~4 data (which shows no variability
over 8 mmag) and, even though the instrumental effects present in the
data were smaller for this object, some of them could still remain
after the corrections.  2MA0908 did not show any rotational modulation
in its light curve and, no other type of variability is present
either.  Hence, we found no variability with limits of 3 mmag and 9
mmag in Ch.~2 and Ch.~4 respectively.

If we assume that the DENIS0255 and 2MA2244 are variable, our simple
model puts an upper limit on the size of the feature in $\sim$1-2\% of
the visible hemisphere of the object. If instead, the variability
shown by our targets has an instrumental origin, our non-variable L
dwarfs could be either completely covered with clouds or objects whose
clouds are smaller and uniformly distributed along its atmosphere.
Such scenarios would lead to very small photometric variations.
Followup photometry in IRAC Ch.~1 and Ch.~3 should distinguish between
instrumental and intrinsic sources of variability.  If the variations
arise on the targets, then the amplitude of the variations should vary
between bandpasses in a manner consistent with the atmospheric
condensate structure \citep{Ackerman01} and still be consistent with
the rotational period implied by our observations.


\acknowledgements
We acknowledge use of the L and T dwarf archives at
http//DwarfArchives.org, maintained by two of us (J.D.K. and C.R.G.)
and Adam Burgasser.  This work is based on observations made with the
$Spitzer$ Space Telescope, which is operated by the Jet Propulsion
Laboratory, California Institute of Technology under a contract with
NASA. Support for this work was provided by NASA through an award
issued by JPL/Caltech.  M.M-C. also acknowledges the funding provided
by the Spitzer Visiting Graduate Students Fellowship Program.

{\it Facilities:} \facility{Spitzer (IRAC)}


\clearpage
\input{tab1.tex}
\input{tab2.tex}
\input{tab3.tex}
\input{tab4.tex}

\clearpage

\end{document}

%% file: tab1.tex
\begin{table}
\small
\caption{Late-L Dwarfs Targets. \label{targets}}
\begin{tabular}{lccccc}
\\
\tableline
\tableline
Object & Opt. Sp. Type & Near-IR Sp. Type & $J-K_s$\tablenotemark{a} &
$v\sin i$ (km s$^{-1}$) & Refs.\tablenotemark{b}\\
\tableline
DENIS0255 & L8  & ---                            & 1.69 \begin{math}\pm\end{math} 0.050 & 40 \begin{math}\pm\end{math} 10 & 1, 5\\
2MA0908   & L5  & L9\begin{math}\pm\end{math}1   & 1.60\begin{math}\pm\end{math}0.051 & 31                                & 2, 4, 6\\
2MA2244   & L6.5& L7.5\begin{math}\pm\end{math}2 & 2.45\begin{math}\pm\end{math}0.213 & ---                               & 3, 4\\
\tableline                                                               
\end{tabular}  
\tablenotetext{a}{$J - K_s$ colors come from the 2MASS magnitudes.}
\tablenotetext{b}{Reference numbers: (1)\cite{Kirkpatrickinprep}
(2)\cite{Cruz03} (3)\cite{Dahn02}  (4)\cite{Knapp04} 
(5)\cite{Basri00} (6)Charbonneau (personal communication 2006)} 
\end{table}

%% file: tab2.tex
\begin{table}
\caption{Observing Strategy.\label{Observations}}
\begin{tabular}{lccccc}
\\
\tableline
\tableline
Object & Obs. Date & $\#$ of & $\#$ of & Frametime & Time on     \\
       & (UT)      & AORs    & Repeats & (sec)     & target (hrs)\\
\tableline
DENIS0255 & 24 Aug 2005 & 6 & 255 &  12 & 6   \\
2MA0908   & 29 Nov 2005 & 1 & 890 &  30 & 7.7 \\
2MA2244   & 29 Nov 2005 & 2 & 104 & 100 & 6.5 \\
\tableline
\end{tabular} 
\end{table}

%% file: tab3.tex
\begin{table}
\caption{Zero magnitude flux for IRAC.\label{calibration}}
\begin{tabular}{ccc}
\\
\tableline
\tableline
IRAC Ch./Wavelength ($\micron$\ )& Flux at zero mag (Jy) & Zero-point magnitude (mag)\\
\tableline
Ch.~2/4.5    & 179.7 & 16.78\\
Ch.~4/8.0    & 64.1  & 15.65\\
\tableline
\end{tabular}
\end{table}

%% file: tab4.tex

\begin{table}
\small
\caption[]{Main results for the 3 targets.\label{results}}
\begin{tabular}{lcccccccc}
\\
\tableline
\tableline
       & \multicolumn{4}{c}{4.5~$\mu$m} &  \multicolumn{4}{c}{8~$\mu$m} \\
Object & \multicolumn{4}{c}{---------------------------------------------------} &  \multicolumn{4}{c}{---------------------------------------------------}\\
          & mag            & RMS   & p     & $ T_{rot}$ (hr) & mag           & RMS   & p   & $T_{rot}$ (hr)\\
\tableline
DENIS0255 & 10.156$\pm$0.002 & 0.006 & $<10^{-4}$ & 7.4                           & 9.519$\pm$0.004 & 0.004 & 0.3 & --- \\
2MA0908   & 11.602$\pm$0.003 & 0.003 & 0.07  & ---                           & 11.067$\pm$0.009& 0.009 & 0.22& --- \\
2MA2244   & 12.083$\pm$0.004 & 0.004 & 0.003 & 4.6                           & 11.346$\pm$0.006& 0.006 & 0.4 & --- \\
\tableline
\end{tabular} 
\end{table}